\documentclass[english,prl,twocolumn,superscriptaddress,amsmath,amssymb,a4paper]{revtex4-1}
\usepackage[T1]{fontenc}
\usepackage[latin1]{inputenc}
\usepackage{graphicx}
\usepackage[top=0.85in,bottom=0.9in,,left=0.75in,right=0.75in]{geometry}
\usepackage{booktabs}
\usepackage{color}
\usepackage{subfigure}
\usepackage{appendix}

\makeatletter
\usepackage{dcolumn}
\usepackage{bm}
\usepackage{epsfig}

\newcommand{\ignore}[1]{}

\usepackage{babel}
\makeatother

\usepackage{sidecap}
\sidecaptionvpos{figure}{c}

\begin{document}

\title{Doping dependence of electronic structure of infinite-layer NdNiO$_2$}

\author{Zhao Liu}
\affiliation{Hefei National Laboratory for Physical Sciences at the Microscale,
University of Science and Technology of China, Hefei, Anhui 230026, China}
\affiliation{Westlake Institution of Advanced Study, Westlake University, Hangzhou 300024, China}

\author{Chenchao Xu}
\affiliation{Westlake Institution of Advanced Study, Westlake University, Hangzhou 300024, China}
\affiliation{Hangzhou Key Laboratory of Quantum Matter, Hangzhou Normal University, Hangzhou 310036, China}

\author{Chao Cao} \thanks{E-mail: ccao@hznu.edu.cn}
\affiliation{Hangzhou Key Laboratory of Quantum Matter, Hangzhou Normal University, Hangzhou 310036, China}

\author{W. Zhu} 
\affiliation{Westlake Institution of Advanced Study, Westlake University, Hangzhou 300024, China}

\author{Z. F. Wang} 
\affiliation{Hefei National Laboratory for Physical Sciences at the Microscale,
University of Science and Technology of China, Hefei, Anhui 230026, China}

\author{Jinlong Yang} \thanks{E-mail: jlyang@ustc.edu.cn}
\affiliation{Hefei National Laboratory for Physical Sciences at the Microscale,
University of Science and Technology of China, Hefei, Anhui 230026, China}


\begin{abstract}
We investigate the electronic structure of nickelate superconductor NdNiO$_2$ upon hole doping, 
by means of density-functional theory and dynamical mean-field theory.
We demonstrate the strong intrinsic hybridization between strongly correlated states formed by Ni-3$d_{x^2-y^2}$ orbital and 
itinerant electrons due to Nd-5$d$ and Ni-3$d_{z^2}$ orbitals, 
producing a valence-fluctuating correlated metal as the normal state of hole-doped NdNiO$_2$.
The Hund's rule appears to play a dominating role on multi-orbital physics in the lightly doped compound, 
while its effect is gradually reduced by increasing the doping level.
Crucially, the hole-doping leads to intricate effects on Ni-3d orbitals,
such as a non-monotonic change of electron occupation in lightly doped level, 
and a flipping orbital configuration in the overdoped regime.   
Additionally, we also map out the topology of Fermi surface at different doping levels.
These findings render a preferred window to peek into electron pairing and superconductivity. 
\end{abstract}

\maketitle

\textit{Introduction.---}Recent discovery of superconductivity in hole-doped infinite-layer NdNiO$_2$
adds the nickelate material to the  family of unconventional superconductivity \cite{nickelate0}. 
While the crystal structure of NdNiO$_2$ is similar to the infinite-layer copper oxide superconductors, 
the electronic structure and magnetic properties of NdNiO$_2$ are very different, which triggers intense on-going discussion. 
Especially, the O-$2p$ states are deeply below the Fermi-level, 
which points to a Mott insulator for NdNiO$_2$ \cite{DFT0, DFT1, DFT2, DFT3}.
The strong insulating behavior is suppressed by the self-doping effect due to hybridization between Nd$-5d$ conduction electrons
and Ni$-3d$ local orbitals \cite{selfdoping1, selfdoping2}, producing a bad metal observed in transport measurements.
Additionally, the related magnetic properties can be understood within the similar route \cite{afm1, afm2, afm3, afm4, afm5}.
Besides these studies on the parent compound,
the microscopic origin for superconductivity is still elusive \cite{mech1, mech2, mech3, mech4, mech5, mech6, mech7}. 
Since the nature of normal state may be responsible for electron pairing and superconductivity, 
it is important to determine their charge doping evolution and connection with superconductivity \cite{dome1, dome2, gapsym}.

Let us recall the effect of hole doping on the well-known  
 copper oxide superconductors where Cu$-3d$ local orbitals is mediated by the O$-2p$ orbitals.
A novel consequence is, upon holes doping most of holes reside on the O$-2p$ orbitals  \cite{cuprate1}. 
The holes in O$-2p$ orbitals will largely weaken the anti-ferromagnetic ordering and induce strong magnetic fluctuations,
driving the system into ${d}$-wave superconductivity \cite{RVB, CC1}. 
Moreover, since the spins on O$-2p$ orbitals strongly interact with local spin on Cu$-3d$ states and form the Cu-O spin singlet (Zhang-Rice state) \cite{Zhang-Rice}, 
a single-band $t$-$J$ model is believed to capture essential physics in cuprates \cite{tJ}.
Since the dynamics of holes play an important role in possible mechanism for superconductivity in cuprates,
it immediately rises questions: 
Does the hole-doping play a similar role in NdNiO$_2$? If not,
what is the nature of normal state of hole-doped NdNiO$_2$?

\begin{figure*}
\centering
\includegraphics[width=17cm]{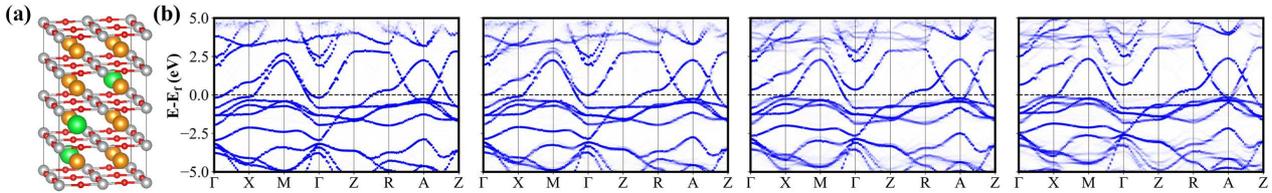}
\caption{(a) Perspective view of NdNiO$_2$ supercell with $\delta$ = 18.75 \% . The silver, red, yellow and green balls represent Ni, O, Nd and Sr atoms, respectively. (b) Unfolded band structures for different Sr doping concentration. 
From left to right: $\delta$ = 0.00 \%, 6.25 \%, 18.75 \% and 31.25 \%. \label{fig:dft_bs} }
\end{figure*}

To address these questions, 
one needs the electron occupation and spin information on multi-orbitals using unbiased computations.
Despite extensive investigations \cite{Ku, Karp, Mei, Olevano, Lech1, Lech2, Chen, Zhang, Zhong2, Held2020}, many open issues still remains, waiting for critical justification.
For example, the hole doping is usually simulated by the shifts of the energy bands \cite{Gao},
nevertheless the possible electronic state induced by doping Sr has not been well addressed. 
Moreover, to avoid the discussion of $f$-electrons on Neodymium \cite{fele}, 
most of existing literatures made a detour and studied LaNiO$_2$ alternatively,
but these calculations are at odds with the experimental fact that
no superconductivity is found in LaNiO$_2$ \cite{nickelate0}. 
In a word, a full-electron \textit{ab initio} simulation of hole-doped NdNiO$_2$ is still outstanding. 

In this work, we aim at the normal state of NdNiO$_2$ by performing both density functional theory (DFT) and  
 dynamical mean field theory (DMFT) calculations. 
Firstly, we build a realistic doping model to study the effect of Sr doping continuously and find that the evolution of the electronic states upon hole doping is non-rigid-band like: 
the electron pocket at $\Gamma$ point vanishes quickly, and the Fermi level is almost pinned by the Ni-$3d_{x^2-y^2}$ orbital.
Then we perform a many-body calculation on the normal state of hole-doped NdNiO$_2$ without considering the $f$-orbitals, and compare with the results from full electron many-body calculations by explicitly considering Nd-4$f$ electrons.
We identify several key features in the evolution of electronic structure upon hole doping:
1) While $d^9$ state is the most probable, the weight of $d^8$ is much larger than $d^{10}$\underline{L} (\underline{L} means a hole on ligand), 
which points to a multi-orbital physics rather than the charge-transfer picture; 
2) The total occupation of Ni-3$d$ state is close to 8.6, thus lightly doped compound is a valence-fluctuating correlated metal; 
3) The Hund's rule plays an important role in the lightly doped compound,
and its effect is gradually reduced upon hole doping; 
4) There is a non-monotonic doping effect in Ni-$3d$ orbital configuration;
5) Significant band renormalization is observed for the Ni-3$d_{x^2-y^2}$ state, 
with mass enhancement of $\sim$ 2.4. It is quickly suppressed to $\sim$ 2.1 upon heavy hole-doping.
We believe these findings are helpful to understand the unconventional superconductivity in this system. 
For instance, the correlated Ni-3$d_{x^2-y^2}$ orbital is most relevant for electron pairing,
and the change of topology of Fermi surface could result in disappearance of superconductivity.

\textit{Methods.---}
DFT calculation was performed with the plane wave projector augmented wave method implemented in the Vienna \textit{ab initio} simulations package (VASP) \cite{vas1,vas2,vas3}. 
The Perdew-Burke-Ernzerhof version of generalized gradient approximation (PBE) was applied \cite{vas4}.  
To simulate the effect of Sr doping, 1, 3 and 5 Nd atoms are replaced by Sr atoms in a ${2 \times 2 \times 4}$ supercell containing 16 Nd, 16 Ni and 32 O atoms, as depicted in Fig. \ref{fig:dft_bs}(a). In this way, we can consistently mimic the hole doping concentration ($\delta$) at 6.25 \%, 18.75\% and 31.25\% lying at the underdoped, 
optimal doping and overdoped regions \cite{dome1, dome2}. 
Considering the fact that the thin films are grown on SrTiO$_3$ (001) surface, the lattice constants in the ab plane of supercell cell are fixed to $a$ = $b$ = 3.92 {\AA} while the lattice constant in the $c$ direction and atomic positions are allowed to relax.  
For band structure calculation, we choose the following high symmetric path $\Gamma$(0, 0, 0)-X($\pi$, 0, 0)-M($\pi$, $\pi$, 0)-$\Gamma$-Z(0, 0, $\pi$)-R($\pi$, 0, $\pi$)-A($\pi$, $\pi$, $\pi$)-Z (see Fig. \ref{fig:dft_bs}(a)). 
To show the effect of Sr doping on electronic structures, all the band structures are  unfolded to the unit cell by PyProcar package \cite{pypro}.

DFT + DMFT calculations were performed using the code EDMFTF, developed by Haule et al. \cite{edmft} based on the Wien2k package \cite{wien2k}.  
Similar to prior studies, we replace Nd with La in our calculations, and the doping effect is simulated by removing conduction electrons in the spirit of virtual crystal approximation (VCA) \cite{VCA}. Charge self-consistency is enforced throughout the calculations. The on-site Coulomb interaction parameters for Ni-3$d$ orbitals are chosen to be $F^0=6.0$ eV, $F^2=7.754$ eV, and $F^4=4.846$ eV, respectively, or equivalently $U=6.0$ eV and $J=0.9$ eV. 
The local impurity problem is solved using continuous-time quantum Monte Carlo (CTQMC) method \cite{cMC} at $\beta=100$ eV$^{-1}$ or 116K. 
For each impurity problem, we used $4\times 10^9$ total CTQMC steps in each DMFT iteration, 
and self energies from last 5 iterations after convergence were averaged and analytically continued to obtain real-frequency self energy $\Sigma(\omega)$ using maximum entropy method. 

\textit{DFT results.---}
We first calculate the lattice constant in the $c$ direction (see Fig. S1), 
and observe a linear relation between Sr doping concentration and  $c$, 
in accordance with experimental result \cite{dome1}. 
After structural optimization, NiO$_2$ plane is no longer exactly flat as in the pristine compound. 
To measure the roughness of the plane, we define roughness as
$ R = \mathrm{max}\{\delta_c \}$,             
where ${\delta_c}$ is the buckling of each NiO$_2$ plane in the supercell (in unit of {\AA}). 
As displayed in Fig. S1, $R$ is less than 0.03 {\AA} even at large $\delta$ = 31.25 \%,  therefore the effect of structure distortion on electronic structure is secondary,
making possible to prepare atomically flat surface for scanning tunneling microscope \cite{gapsym}.

\begin{figure*}
\centering
\includegraphics[width=15cm]{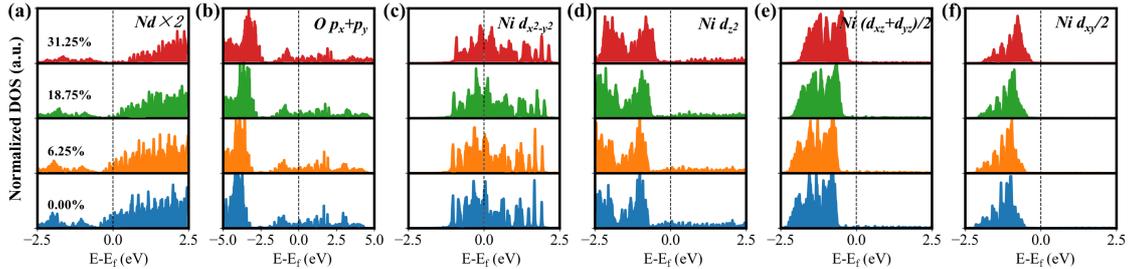}
\caption{(a)-(d) Projected density of state with different $\delta$ on different subspaces, from left to right: Nd atoms, O $p_{x+y}$ orbitals, Ni $d_{x^2-y^2}$ orbitals, Ni $d_{z^2}$ orbitals, Ni $d_{xz}+d_{yz}$ orbitals and Ni $d_{xy}$ orbitals. The weight in (a), (e) and (f) has been rescaled by a factor of 2, 1/2 and 1/2. From bottom to top, $\delta$ increases from 0.00 \% to 31.25 \% in each figure. \label{fig:dft_dos}}
\end{figure*}

The unfolded band structures at different $\delta$ are shown in Fig. \ref{fig:dft_bs}(b), 
where the band structure of pristine compound is also plotted to guide the eye. To facilitate our discussion, 
we label the Fermi pockets as $\alpha$, $\beta$ and $\gamma$ for the dominating sheet, 
pocket at $\Gamma$ point and pocket at A point in the pristine compound \cite{afm1, gapsym}. 
These pockets behave differently to Sr doping. Among these Fermi pockets, $\beta$ pocket is very sensitive to Sr doping and almost vanishing at a low doping level $\delta$ = 6.25 \%. 
Since it is contributed by Nd $d_{z^2}$ and Ni $d_{z^2}$ orbitals, 
their density of states (DOS) gradually decrease to zero as shown in Fig. \ref{fig:dft_dos}(a) and \ref{fig:dft_dos}(d). 
When $\delta$ increases to 31.25 \%, the $\beta$ pocket has experienced a blue-shift up to 0.86 eV. 
Such a large shift indicates the free-particle nature of this pocket. The $\gamma$ pocket, 
which has smaller effective electron mass and thus more itinerant, nevertheless has smaller blue-shift than $\beta$ pocket and does not vanish even at $\delta$ = 31.25 \%. 
In a free-particle picture, a homogeneous holes doping will lead to the drop of Fermi level, namely, 
all the bands are rigid and have the same blue-shift. The smaller blue-shift of the more itinerant $\gamma$ pocket reflects that such a pocket is actually more correlated. 
As we will show below, such a correlation originates from the Hund's coupling between Ni-3$d$ for this pocket is mainly formed by the Ni $d_{xz}$ and $d_{yz}$ orbitals \cite{afm1}.
As for the $\alpha$ sheet, its blue-shift strongly depends on $\delta$. At smaller $\delta$, it does not shift until the vanishing of $\beta$ pocket(see Fig. \ref{fig:dft_dos}(c)). 
After that, it slowly undergoes blue-shift together with $\gamma$ pocket.  Moreover, since $\alpha$ sheet wraps around a large volume, its blue-shift is quite small. In other words, the Fermi level is pinned to $\alpha$ sheet. Consider the fact that $\alpha$ sheet are mainly contributed by Ni $d_{x^2-y^2}$ orbital, such a Fermi surface pinning by Ni $d_{x^2-y^2}$ orbital leads to two results:
1) the charge transfer energy between Ni $d_{x^2-y^2}$ and O-$2p$ is reduced since O-$2p$ experiences a large blue-shift (as shown in Fig. \ref{fig:dft_dos}(b)) \cite{Kirshna2020}. 
Such a decrease in charge transfer energy in NiO$_2$ plane indicates more contributions from O-$2p$ orbitals in the Fermi level, which may explain the existence of "shoulder" structure in the O-K edge resolved by electron energy loss spectroscopy (EELS) recently \cite{eels}.
2) the splitting between Ni $d_{x^2-y^2}$ and other Ni $d$ orbitals becomes smaller, which drives NdNiO$_2$ towards multi-orbital Hund's physics \cite{Hund}. 

We note that the above features upon doping are quite robust since similar behavior has been repeated with SCAN functional (see SM). 
In addition, vanishingly small DOS of Sr is identified around the Fermi level, therefore Sr doping does not introduce any Sr orbitals near the Fermi level. 
Thus it is valid to ignore the chemical differences by removing valence electrons while simulating the doping effect.

\begin{figure*}
\includegraphics[width=16cm]{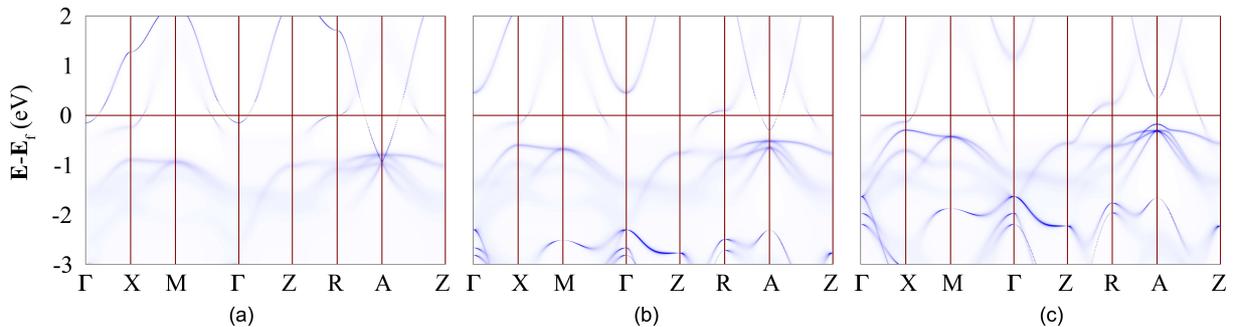}
\caption{$\mathbf{k}$-resolved spectral function from DFT+DMFT calculations for (a) pristine, (b) 0.2 hole-doped, and (c) 0.4 hole-doped NdNiO$_2$ at 116 K. 
\label{fig:dmft_spec}}
\end{figure*}

\textit{DFT + DMFT results.---}
We now turn to many-body results from DFT+DMFT calculations. In Fig. \ref{fig:dmft_spec}, we present the $\mathbf{k}$-resolved spectral function for pristine, 0.2 hole doped, and 0.4 hole doped NdNiO$_2$ from DFT+DMFT calculations. Despite of the relatively high temperature (116K), the electron states in the pristine compound are already very coherent near the Fermi level, in accordance with previous reports \cite{ryee, afm2}. The spectral function of the pristine compound is quite similar to the DFT band structure, giving rise to three sheets of Fermi surfaces with similar shapes. Hole-doping quickly suppresses the $\beta$ pocket, which disappears with less than 0.1 hole doping. Therefore, this Nd-$d_{z^2}$/Ni-$d_{z^2}$ derived pocket may be irrelevant to the superconductivity. The $\gamma$ pocket is also substantially reduced by hole-doping, and eventually disappears around 0.3 hole-doping. The largest $\alpha$ pocket is only moderately affected before 0.3 hole-doping, due to its large volume and partial flat-band close to the Fermi level at R. Further hole doping also causes substantial blue-shift of $\alpha$ pocket. We show the calculated DMFT Fermi surfaces in Fig. \ref{fig:dmft_fs}. In the pristine compound, the occupation of $\alpha$, $\beta$, and $\gamma$ pockets are 0.98 holes, 0.03 electrons, and 0.02 electrons, respectively. Upon 0.2 hole doping, the occupation of $\alpha$ and $\gamma$ pocket are 1.19 holes and $<$0.01 electrons, respectively, while the $\beta$ pocket disappears. In the 0.4 hole doped compound, only $\alpha$ pocket remains with 1.32 hole occupation. Such observation is also consistent with the above DFT analysis, and follows Luttinger theorem. 
Moreover, as a result of the large blue-shift of $\beta$ and $\gamma$ pockets upon hole doping, as well as the pinning of the $\alpha$-pocket due to Ni-3$d_{x^2-y^2}$, the hybridization between the itinerant electrons and local Ni$-3d$ orbitals reduces, resulting in sharper quasi-particle behavior upon hole-doping [comparing Fig. \ref{fig:dmft_spec}(b,c) with (a) (see also Fig. \ref{fig:hyb})].

\begin{figure*}
\includegraphics[width=16cm]{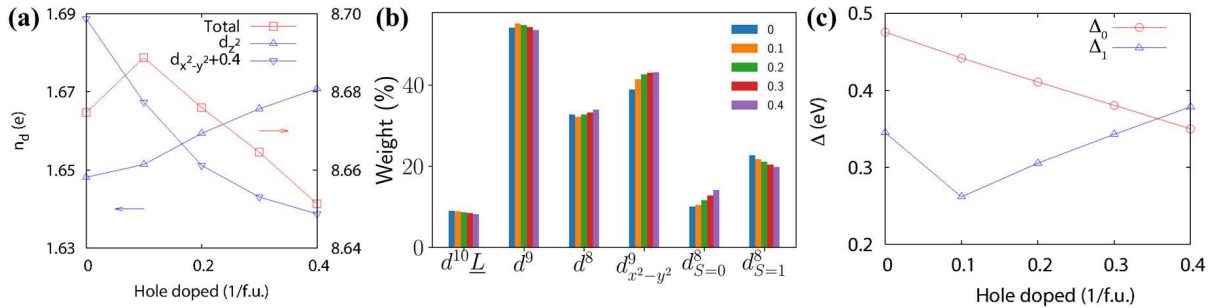}
\caption{Evolution of (a) Ni-3$d$ $e_g$ orbital occupation, (b) weight of many-body states, and (c) Ni-3$d$ orbital crystal field splitting under hole doping. In panel (a), the occupation of $d_{x^2-y^2}$ is increased by 0.4 to fit in the plot. In panel (b), $d^9_{x^2-y^2}$ indicates $d^9$ state with unpaired $d_{x^2-y^2}$ electron, $d^8_{S=0}$/$d^8_{S=1}$ are $d^8$ spin-singlet/spin-triplet states, respectively. In panel (c), $\Delta_0$ ($\Delta_1$) are the energy differences between $d_{x^2-y^2}$ ($d_{zx/zy}$) and $d_{xy}$ orbitals, respectively.
\label{fig:occ_cef}}
\end{figure*}

We then focus on the 3$d$ orbital occupations of Ni atom (Fig. \ref{fig:occ_cef}(a), also TAB. \ref{tab:occ}). The total Ni-3$d$ occupation in the pristine compound is 8.675 $e$, thus the Ni atom is valence fluctuating between Ni$^{2+}$ ($d^8$) and Ni$^{+}$ ($d^9$). Despite of the consistent and significant changes in the spectral function due to hole doping, the doping effect to orbital occupations clearly exhibits different behavior for small and large doping range. Surprisingly, for small hole doping ($<0.1$ hole/cell), the overall Ni-3$d$ occupation even increases. Once the hole-doping exceeds 0.1 hole/cell, the overall Ni-3$d$ occupations follow a consistent linear decreasing behavior with respect to the hole-doping level. Closer examination reveals further details. Among the 5 Ni-3$d$ orbitals, the $t_{2g}$ orbitals are nearly fully occupied, and are less relevant to the electronic states near the Fermi level.
Therefore, the $t_{2g}$ orbitals are "frozen" during hole doping, although they cannot be ignored in the discussion of total $d$-occupation. 
In the following discussion, we focus on the $e_{g}$ orbitals \cite{Karp, Held2020, Lech2, afm5}. 
The doped hole will populate on $d_{x^2-y^2}$, as the occupation decreases with hole doping as shown in Fig. \ref{fig:occ_cef}(a). 
Interestingly,  $d_{z^2}$ occupation increases with hole doping. 
These findings are at odds with Ref. \cite{Lech1}, where the hole mainly goes to $d_{z^2}$ orbital.
Such a charge correlation between $d_{z^2}$ and $d_{x^2-y^2}$ agrees with the band shift upon hole doping from DFT calculations. In addition, the hole doping has nontrivial effect on the effective mass of the Ni-3$d$ orbitals as well. As shown in Tab. \ref{tab:occ}, Ni-3$d_{x^2-y^2}$ orbital has the largest $m^*/m\sim2.4$, and all other $d$ orbitals show only mild correlation effect with $m^*/m$ around 1.3 in the pristine compound. While the later is less affected by hole doping, the effective mass of $d_{x^2-y^2}$, which is believed to be most relevant to the superconductivity, strongly depends on the doping. It ranges between 2.4 to 2.6 within 0.3 hole/cell doping, and suddenly drops to $\sim$2.1 with 0.4 hole/cell doping.

We also plot the statistical weight of sampled many-body states in Fig. 4(b). Again, we see clearly distinct hole-doping behavior for the small and large doping range. 
For the small doping range ($<0.1$ hole/cell), the $d^9$ weight is significantly enhanced, primarily due to large increase of $d^9_{x^2-y^2}$ weight. Noticing that $d^9_{x^2-y^2}$ denotes $d^9$ with unpaired $d_{x^2-y^2}$, this is consistent with the reduced $d_{x^2-y^2}$ occupation due to hole doping.
Beyond 0.1 hole/cell doping, the $d^9$ weight consistently drops, whereas the $d^9_{x^2-y^2}$ weight gradually reaches saturation around 43\% at 0.4 hole/cell. Similarly, the $d^8$ configuration weight reduces from 32.7\% to 32.1\% in the small doping range, then linearly increases to 33.9\% at 0.4 hole/cell doping. Nevertheless, within the doping range we investigated, the weight of $d^8$ configuration is always more than 3-times larger than $d^{10}$\underline{L}, and even enhances with 0.4 hole-doping. In a typical charge-transfer insulator, it is the weights of $d^{10}$\underline{L} that should be comparable to that of $d^9$ rather than $d^8$ \cite{Karp}. Thus it suggests the multi-orbital nature rather than charge-transfer. Moreover, in the pristine and small doping cases, the spin-triplet states always dominate the $d^8$ states (almost twice the weight of spin-singlet states). However, the weight ratio between spin-triplet and singlet states quickly reduces, and becomes close to 1 in the 0.4 hole/cell doped case. Therefore, the Hund's physics is important in the pristine and lightly-doped system, but may lose its importance in the heavily doped compounds.

Such non-monotonic doping effect is also observed for the crystal-field splitting of the Ni-3$d$ electrons. Bare atomic level of Ni-3d orbitals (with respect to the Fermi level) can be obtained by fitting the undoped DFT result to a tight-binding Wannier Hamiltonian, which yields -1.398 eV, -1.459 eV, -1.676 eV, and -1.919 eV for $d_{x^2-y^2}$, $d_{z^2}$, $d_{zx/zy}$, $d_{xy}$ orbitals, respectively. 
The $e_{g}$ orbitals are nearly degenerate and is separated from $t_{2g}$ by a large gap. Such crystal field splitting is consistent with the fact that the $t_{2g}$ are "frozen" during hole doping.
In the DFT+DMFT calculations, the electron density and energies are renormalized, and the orbitals order as $E(d_{z^2})$ > $E(d_{x^2-y^2})$> $E(d_{zx/zy})$ > $E(d_{xy})$ in the pristine compound. 
Since $d_{z^2}$ has larger occupation than $d_{x^2-y^2}$, the orbital order flip between $d_{z^2}$ and $d_{x^2-y^2}$ reflects the fact that $d_{x^2-y^2}$ has much larger onsite Hubbard repulsion than $d_{z^2}$, which renders the validity of the one-band Hubbard model\cite{Held2020}. Since $d_{x^2-y^2}$ occupation is larger than half-filled, the residue of hole on $d_{x^2-y^2}$ will greatly stabilize the system as found in Fig. \ref{fig:occ_cef}a. The observation of stronger correlation of $d_{x^2-y^2}$ than $d_{z^2}$ is also in agreement with the fact that $d_{x^2-y^2}$ has a much larger effect mass than $d_{z^2}$.
Normally, for D$_{4h}$ symmetry without apical anions, crystal field splitting leads to lowest $d_{zx/zy},d_{z^2}$ instead of $d_{xy}$ orbitals. However, the NdNiO$_2$ compound stems from NdNiO$_3$ crystal, where the octahedral crystal-field leads to degenerate $t_{2g}$ and $e_g$ splittings for Ni-3$d$ orbitals. Once the apical O$^{2-}$ anions are removed, negative charge occupation accumulates at the apical oxygen position\cite{DFT2, Chen}, and thus orbitals extending along the $z$-direction costs energy, raising energies of $d_{z^2}$, $d_{zx}$ and $d_{zy}$ orbitals. Therefore, the $d_{xy}$-orbital becomes the lowest-lying component. The $d_{zx/zy}$ orbitals are less affected than the $d_{z^2}$, and therefore are also low-lying and nearly fully occupied. In such distorted octahedral crystal field, the original octahedral crystal-field splitting is characterized by the energy difference between $d_{x^2-y^2}$ and $d_{xy}$ orbitals $\Delta_0=E(d_{x^2-y^2})-E(d_{xy})$, and the effect of distortion is reflected in the energy difference between $d_{xy}$ and $d_{zx/zy}$ orbitals $\Delta_1=E(d_{zx/zy})-E(d_{xy})$. We plot the doping dependent $\Delta_0$ and $\Delta_1$ in Fig. \ref{fig:occ_cef}(c). Throughout the hole doping range, $\Delta_0$ exhibits linear dependence on the doping level, but $\Delta_1$ decreases below 0.1 hole doping, and then quickly linearly increases with further hole-doping. This leads to an inversion between $d_{x^2-y^2}$ and $d_{zx/zy}$ at 0.4 hole doping. 

Finally, we address the effect of undetermined double counting term. The above reported results were obtained using a fixed double-counting $V_{dc}=47.4$ eV (or equivalently nominal $n^0=9.0$), which leads to actual $n^d$ close to 8.6. A $n^d=9.0$ solution can be achieved by using $V_{dc}=51.84$ eV. In this case, the $\beta$ pocket disappears even in the pristine compound, and the size of $\gamma$ pocket is drastically reduced compared to the above results. Hole-doping in this case will quickly suppress the $\gamma$ pocket, leaving only $\alpha$ pocket in the BZ. The $d^{10}$\underline{L} configuration weight is much enhanced and can be comparable to the $d^8$ weight (Tab. \ref{tab:stat1}). In this case, the hole doping quickly suppress the effect of Hund's rule, and the weight of spin-singlet states are comparable to the spin-triplet states as early as 0.2 hole/cell doping. Nevertheless, the aforementioned non-monotonic doping effect to the $d$-orbital occupation, many-body configuration weight, crystal-field splitting and effective mass still remains (Fig. \ref{fig:occ_cef1}, Tab. \ref{tab:occ1}, Tab. \ref{tab:stat1}). Thus, our results are robust with respect to the choice of double counting terms. 

\textit{Summary and Discussion.---}
We have performed 
a systematic study of electronic properties of the normal state of NdNiO$_2$ by DFT and DMFT methods.
The complementary methods show  the normal state of hole-doped NdNiO$_2$ as a valence-fluctuating correlated metal.
The Hund's effect and hybridization effect gradually reduces by increasing the doping level. 
Importantly, we identify a non-monotonic many-body doping effect on Ni-$3d$ orbitals,
as evidenced by the electron occupation, orbital crystal field splitting, and 
statistical spectral weight of orbital configurations.
Additionally, the doping effect leads to the change of topology of Fermi surface and reduction of correlations,
which could be relevant to the underlying superconductivity \cite{gapsym}.

We would like make several remarks here. 
First of all, our results show the normal state of doped NdNiO$_2$ is quite different from 
the copper oxide superconductors, instead it is closer to the iron-based superconductors 
where multi-orbital physics dominates. 
Second, we identify  
Ni-$3d_{x^2-y^2}$ orbital as the most correlated one, 
which should be helpful for building the effective model.  
Third, some features shown here could provide plausible understanding for experimental observations. 
For example, both DFT and DFT+DMFT calculations (Fig. \ref{fig:dft_dos} and \ref{fig:dmft_spec}) show O$-2p$ state moving towards Fermi level  upon Sr doping, which 
is consistent with a recent EELS study on O-K edge that gives a  ``shoulder'' structure pointing to Ni-O singlet state \cite{eels}. 
Finally, we argue that the Nd $f$-electrons influence the electronic states below the Fermi energy, leaving the physics close to the Fermi level unaffected. In addition, the occupations and effective masses of Ni-3$d$ orbitals in both pristine and hole-doped compounds are qualitatively similar to the case without $f$-electrons. These arguments are supported by our full-electron calculation of NdNiO$_2$ by explicitly including Nd-4$f$ orbitals (see Supplemental materials Tab. \ref{stab:occf}, FIG. \ref{sfig:dmft_spec}). 
These results all indicates the $f$-electrons may be irrelevant in NdNiO$_2$, 
in good agreement with the existing experiment \cite{Osada2020}.

\textit{Acknowledgement.---}
The authors thanks J.-X. Zhu, Zhi Ren, and H. H. Wen for inspiring and helpful discussions.
This work was supported by NSFC (No. 11774325, 11874137, 21603210, 21603205, 21688102),
National Key Research and Development Program of China (No. 2017YFA0204904, 2016YFA0200604),
Anhui Initiative in Quantum Information Technologies (No. AHY090400),
Fundamental Research Funds for the Central Universities and the Start-up Funding
from Westlake University. We thank Supercomputing Center at USTC for providing
the computing resources. DMFT calculations were performed on the High Performance Computing Cluster at Hangzhou Normal University, as well as at the Beijing Supercloud Computing Center.

\textit{Note added.---} In the final stage of this work, we become aware of 
recent work \cite{Peto2020} using GW+DMFT and \cite{Kotliar1, Kotliar2} using DFT+DMFT. 


\clearpage
\renewcommand\thefigure{\thesection S\arabic{figure}}
\setcounter{figure}{0} 
\renewcommand\thetable{\thesection S-\Roman{table}}
\setcounter{table}{0} 

\appendix

In this supplementary materials, we provide additional results to 
support the discussion in the main text.

\section{I. Structure optimization}
The doping configuration with $\delta$ = 6.25 \%, 18.75 \% and 31.25 \% is shown in Fig. S1(a). Since Sr$^{2+}$ has larger ionic radius than Nd$^{3+}$, the lattice constant in the c direction will expands when Sr is doped. The calculated lattice constant of c direction with respect to $\delta$ is shown in Fig. S1(b), where lattice constant monotonically increases with $\delta$, in accordance with expeimental data \cite{dome}. Since the distortion of NiO$_2$ plane will hugely influence its correlations, we also measure the flatness of each NiO$_2$ plane in the supercell after structural optimization. As shown in Fig. S1(b), the roughness of NiO$_2$ plane is only 0.03 {\AA} at a high doping concentration $\delta$ = 31.25 \%. Therefore, a atomic flat surface is in principle possible.

\begin{figure*}
\centering
\includegraphics[width=17cm]{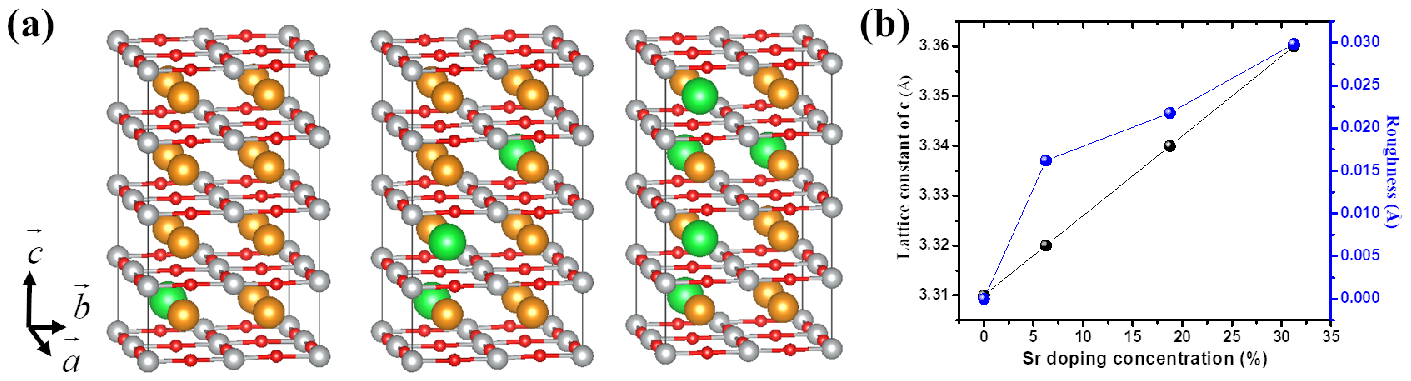}
\caption{(a) From left to right: perspective view of ${\rm NdNiO_2}$ supercell with $\delta$ = 6.25 \%, 18.75 \% and 31.25 \%. The silver, red, yellow and green balls represent Ni, O, Nd and Sr atoms. (b) Lattice constant in c direction and roughness with respect to $\delta$.}
\end{figure*}

\section{II. Result of SCAN functional}
The recently developed SCAN functional \cite{SCAN1} makes it possible to treat charge, spin and lattice degrees of freedom on equal footing. As a parameter-free functional, SCAN has shown great success in different types of bonding systems \cite{SCAN2}, even in cuprates \cite{SCAN3, SCAN4, SCAN5}, which is believed to be the typical strongly-correlated system. Therefore, we also use SCAN functional to study the effect of Sr doping. 

The unfolded band structure at $\delta$ = 0.00 \%, 6.25 \%, 18.75 \% and 31.25 \% is shown in Fig. S2. Similar to the result of PBE, the evolution of electronic structure upon hole doping is non-rigid-like:
1) $\beta$ pocket undergoes Lifshitz transition at a low $\delta$ around 6.25 \%. 
2) $\gamma$ pocket does not vanish even at $\delta$ = 31.25 \%. 
3) the $\alpha$ sheet pins the Fermi level.  
The projected density of state is shown in Fig. S3. In accordance with PBE, SCAN functional gives similar result on the evolution of electronic structure with hole doping. 

\begin{figure*}
\centering
\includegraphics[width=17cm]{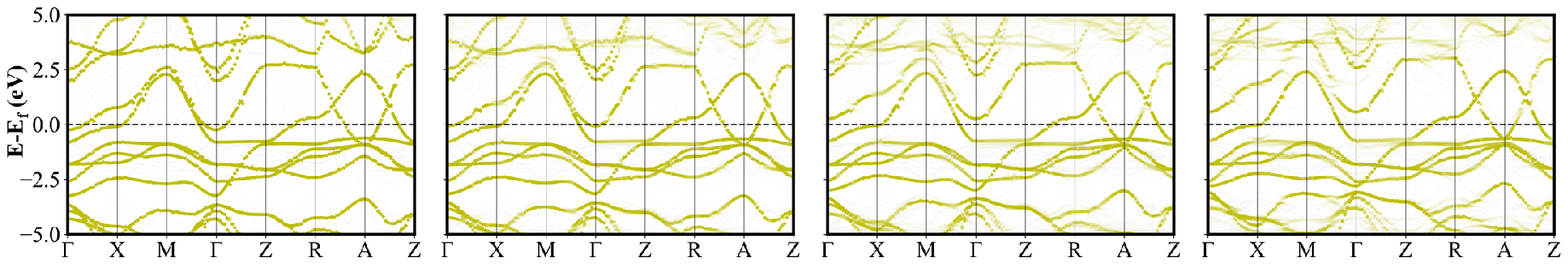}
\caption{Unfolded band structures for different Sr doping concentration. 
From left to right: $\delta$ = 0.00 \%, 6.25 \%, 18.75 \% and 31.25 \%. }
\end{figure*}

\begin{figure*}
\centering
\includegraphics[width=17cm]{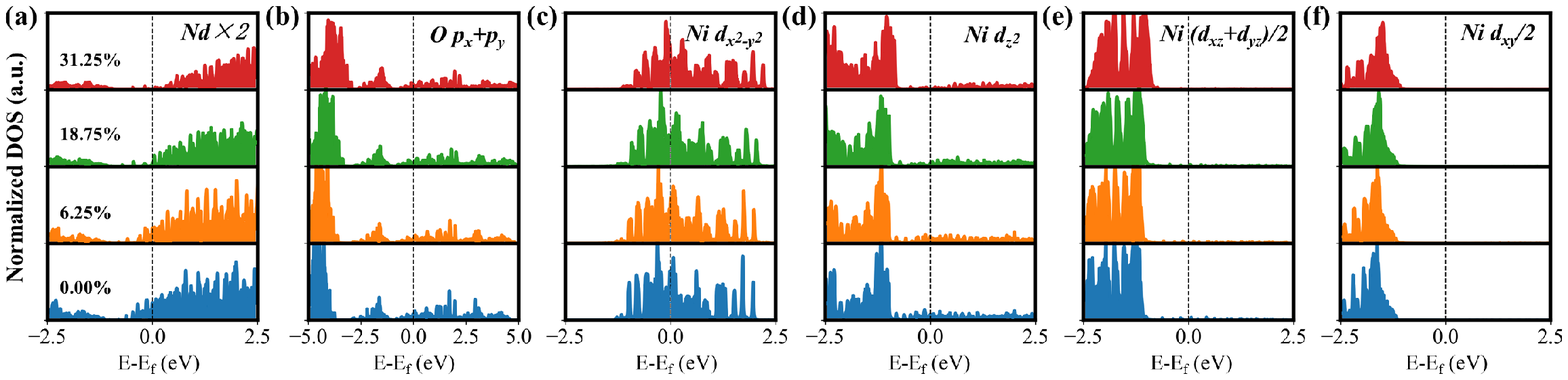}
\caption{(a)-(f) Projected density of state with different $\delta$ on different subspaces, from left to right: Nd atoms, O $p_{x+y}$, Ni $d_{x^2-y^2}$, Ni $d_{z^2}$, Ni $d_{xz}+d_{yz}$ and Ni $d_{xy}$ orbitals. The weight in (a), (e) and (f) has been rescaled by a factor of 2, 1/2 and 1/2. From bottom to top, $\delta$ increases from 0.00 \% to 31.25 \% in each panel. }
\end{figure*}

\section{III. Additional Result of DMFT}

\begin{figure*}
\includegraphics[width=12cm]{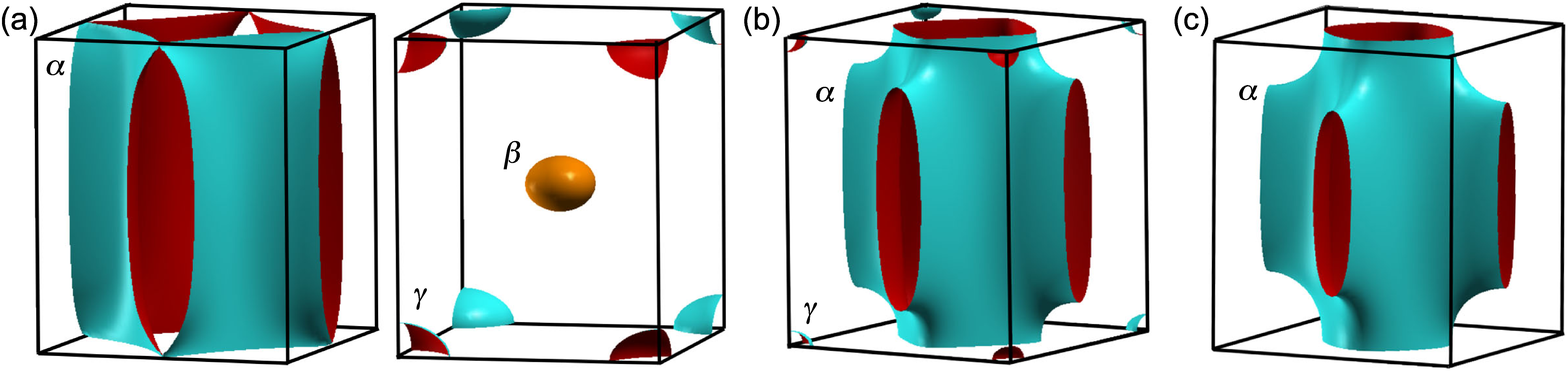}
\caption{DFT+DMFT Fermi surfaces of (a) Pristine NdNiO$_2$, (b) 0.2 hole-doped NdNiO$_2$ and (c) 0.4 hole-doped NdNiO$_2$ from $V_{dc}=47.4$ eV calculations. 
For clarity purposes, the dominating $\alpha$ sheet is separated from $\beta$ and $\gamma$ pockets in panel (a). \label{fig:dmft_fs}}
\end{figure*}

In Fig. \ref{fig:dmft_fs}, we show the DFT+DMFT Fermi surfaces for pristine and doped compounds. The pristine compound consists of three Fermi surface sheets, namely the dominating $\alpha$ pocket due to Ni-3$d_{x^2-y^2}$ orbitals, the electron-type $\beta$ pocket around $\Gamma$, and the electron-type $\gamma$ pocket around R. With 0.2 hole-doping, the $\beta$ pocket disappears, and the $\gamma$ pocket shrinks. Upon 0.4 hole doping, both $\beta$ and $\gamma$ pockets disappears.

\begin{figure*}
\includegraphics[width=12cm]{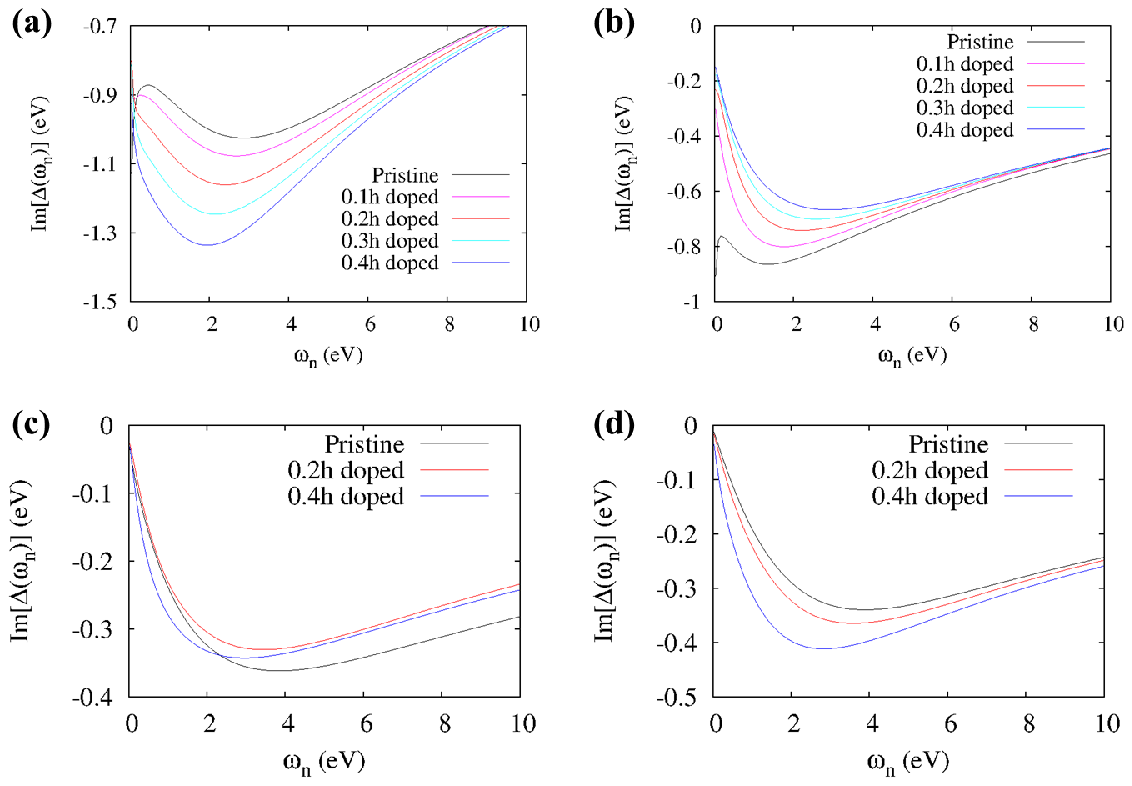}
\caption{Doping dependence of hybridization function for (a) Ni-3$d_{x^2-y^2}$, (b) Ni-3$d_{z^2}$, (c) Ni-3$d_{zx/zy}$ , and (d) Ni-3$d_{xy}$ orbitals in $V_{dc}=47.4$ eV calculations.\label{fig:hyb}}
\end{figure*}

In Fig. \ref{fig:hyb}, we show the hybridization function in Matsubara frequency from DFT+DMFT calculations. In pristine compounds, both Ni-3$d_{x^2-y^2}$ and Ni-3$d_{z^2}$ orbitals show strong hybridization with the conduction electrons, as indicated by their divergent behavior with $\omega_n\rightarrow0$. A small doping with 0.1 hole quickly suppresses the divergent behavior of both orbitals. The doping effect on $t_{2g}$ orbital hybridization functions is much less prominent.

In Tab. \ref{tab:occ} we show the orbital occupation and effective masses in both pristine and doped compounds, calculated with $V_{dc}=47.4$ eV. Decreasing of $d_{x^2-y^2}$ effective mass from 0.1 hole/cell doping to 0.4 hole/cell doping can be observed, and a sudden drop from 0.3 hole/cell doping to 0.4 hole/cell doping can be identified. Similar decreasing of $d_{x^2-y^2}$ effective mass can also be identified in $V_{dc}=51.84$ eV calculations (Tab. \ref{tab:occ1}), however the sudden drop from 0.3 to 0.4 hole/cell doping is absent in $V_{dc}=51.84$ eV calculations. 

\begin{figure*}
\includegraphics[width=12cm]{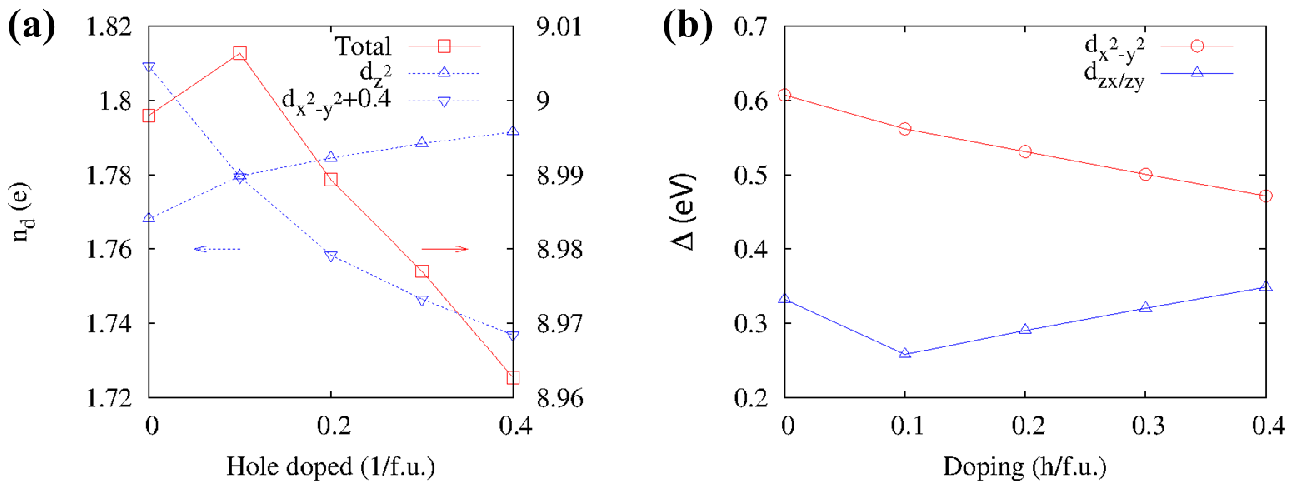}
\caption{Evolution of (a) Ni-3$d$ orbital occupation and (b) Ni-3$d$ orbital crystal field splitting under hole doping, calculated using $V_{dc}=51.84$ eV. \label{fig:occ_cef1}}
\end{figure*}

In Tab. \ref{tab:stat1}, we show the statistic weight of sampled many-body configurations in the pristine and doped compounds for $V_{dc}=51.84$ eV. Although the $d^{10}$\underline{L} configuration weight is substantially higher than $V_{dc}=47.4$ eV calculations, the non-monotonic doping effect can still be observed. Similarly, the non-monotonic doping effect can also be identified from the Ni-3$d$ occupation and crystal field splittings in $V_{dc}=51.84$ eV calculations (Fig. \ref{fig:occ_cef1}).

\begin{table}
  \caption{Ni-3$d$ orbital occupations and effective masses in both pristine and hole doped compounds for $V_{dc}=47.4$ eV. The numbers within the brackets are the effective mass of respective orbitals calculated using $m^*/m=1/Z=1-\frac{\partial\mathrm{Im}[\Sigma(i\omega_n)]}{\partial \omega_n}|_{\omega_n\rightarrow0}$ \label{tab:occ}.}
  \begin{tabular}{c|c|c|c|c}
    \hline\hline
     doping  & $d_{x^2-y^2}$ & $d_{z^2}$ & $d_{zx/zy}$ & $d_{xy}$  \\
    \hline
    0.0 & 1.289 (2.40) & 1.648 (1.26) & 3.797 (1.25) & 1.941 (1.30) \\
    0.1 & 1.267 (2.58) & 1.651 (1.28) & 3.826 (1.27) & 1.944 (1.32) \\
    0.2 & 1.251 (2.43) & 1.659 (1.26) & 3.826 (1.27) & 1.939 (1.32) \\
    0.3 & 1.243 (2.43) & 1.666 (1.29) & 3.824 (1.28) & 1.932 (1.33) \\
    0.4 & 1.239 (2.06) & 1.671 (1.29) & 3.819 (1.29) & 1.923 (1.31) \\
    \hline\hline
  \end{tabular}
\end{table}

\begin{table}
  \caption{Ni-3$d$ orbital occupations and effective masses in both pristine and hole doped compounds for $V_{dc}=51.84$ eV. The numbers within the brackets are the effective mass of respective orbitals. \label{tab:occ1}.}
  \begin{tabular}{c|c|c|c|c}
    \hline\hline
     doping  & $d_{x^2-y^2}$ & $d_{z^2}$ & $d_{zx/zy}$ & $d_{xy}$  \\
    \hline
    0.0 & 1.409 (1.78) & 1.768 (1.20) & 3.860 (1.20) & 1.960 (1.23) \\
    0.1 & 1.379 (1.85) & 1.780 (1.20) & 3.884 (1.20) & 1.963 (1.23) \\
    0.2 & 1.358 (1.84) & 1.785 (1.20) & 3.885 (1.20) & 1.962 (1.24)\\
    0.3 & 1.346 (1.79) & 1.788 (1.20) & 3.883 (1.20) & 1.959 (1.23) \\
    0.4 & 1.337 (1.74) & 1.792 (1.21) & 3.880 (1.20) & 1.954 (1.23) \\
    \hline\hline
  \end{tabular}
\end{table}

\begin{table}
  \caption{Statistic weight of many-body configurations in pristine and doped compounds for $V_{dc}=51.84$ eV.\label{tab:stat1}}
  \begin{tabular}{c|c|c|c|c|c}
    \hline\hline
    doping & $d^{10}$\underline{L} & $d^9$ & $d^9_{x^2-y^2}$ & $d^8_{S=0}$ & $d^8_{S=1}$ \\
    \hline
    0.0 & 21.4\% & 58.5\% & 41.9\% & 6.4\% & 12.1\% \\
    0.1 & 21.2\% & 59.5\% & 45.3\% & 6.8\% & 11.1\% \\
    0.2 & 20.3\% & 59.7\% & 46.9\% & 7.5\% & 11.0\%  \\
    0.3 & 19.8\% & 59.7\% & 47.8\% & 8.1\% & 10.9\% \\
    0.4 & 19.2\% & 59.5\% & 48.3\% & 8.9\% & 10.8\%  \\
    \hline\hline
  \end{tabular}
\end{table}


\section{IV. Crystal field splitting of d orbitals}
\begin{figure*}
\centering
\includegraphics[width=14cm]{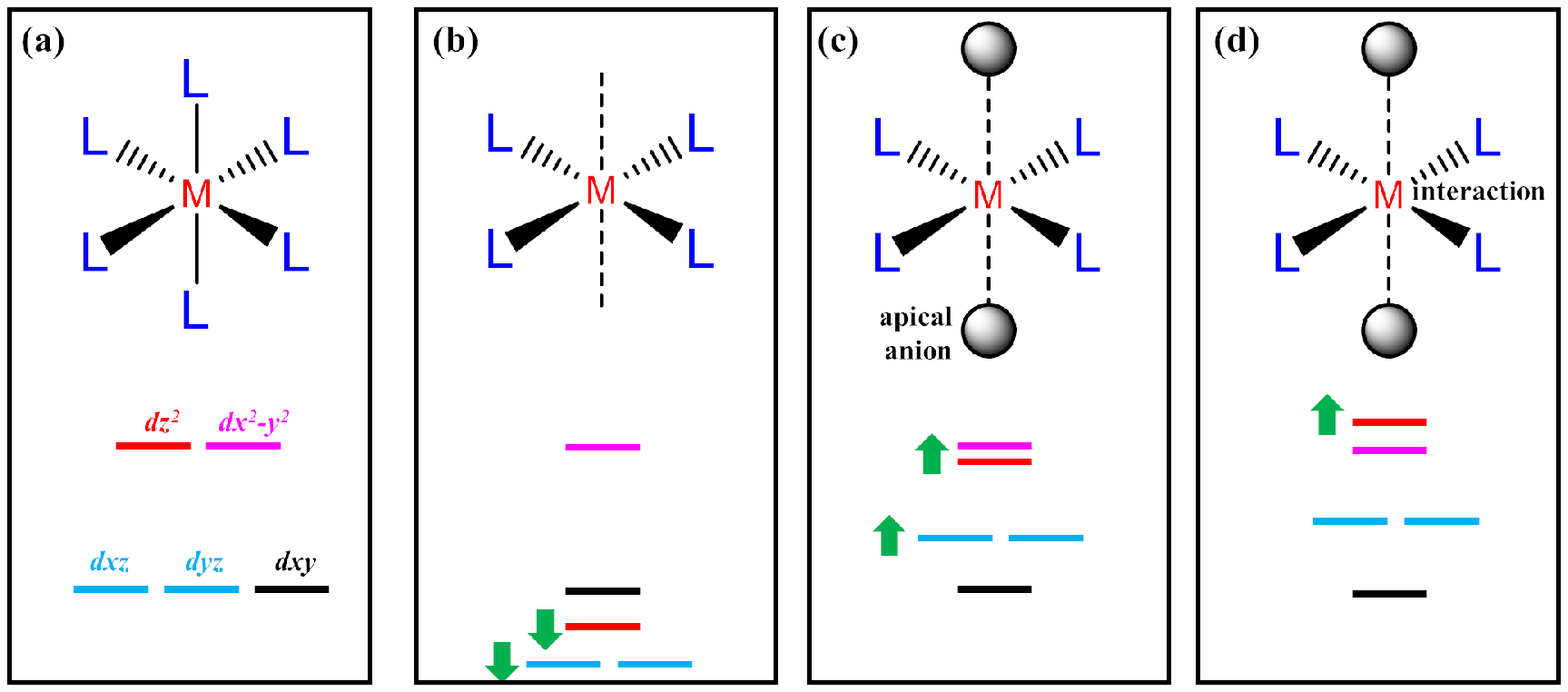}
\caption{Crystal field splitting of d orbitals with (a) ${O_h}$ symmetry (b) ${D_{4h}}$ (c) ${D_{4h}}$ plus apical anions and (d) ${D_{4h}}$ plus apical anions with interaction. The d orbitals are labelled by different colors: $d_{x^2-y^2}$, $d_{z^2}$, $d_{xz}$($d_{yz}$) and $d_{xy}$ orbital corresponds to pink, red, blue and black color.The greed filled arrows indicate how orbital order evolves with new effects.  \label{fig:cef_cartoon}}
\end{figure*}

Here we focus on the crystal field splitting of d orbitals in ${\rm NdNiO_2}$. To begin with, the splitting ${O_h}$ point group is shown in Fig. \ref{fig:cef_cartoon}(a), where {$d_{x^2-y^2}$,  $d_{z^2}$} are two-fold degenerated and
{$d_{xz}$, $d_{yz}$, $d_{xy}$} are three-fold degenerated. The ${D_{4h}}$ point group can be obtained by moving the two vertical ligands to infinity as shown by the dashed line in Fig. \ref{fig:cef_cartoon}(b). Since the d orbitals are anti-bonding type, such lifting of vertical ligands will weak the anti-bonding d orbitals, the result is orbital energy lowering as indicated by the green arrows in Fig. \ref{fig:cef_cartoon}(b). The other two orbitals-$d_{x^2-y^2}$, $d_{xy}$-experience no affect for their planar orbital character. This picture can be modified by the apical anions as shown in Fig. \ref{fig:cef_cartoon}(c), where $d_{z^2}$, $d_{xz}$ and $d_{yz}$ are pushed to higher energy. Therefore, the $d_{xy}$ now has the lowest energy while the $d_{x^2-y^2}$ has the highest energy. Such an orbital order will change with the inclusion of interaction as shown in Fig. \ref{fig:cef_cartoon}(d). The $d_{x^2-y^2}$ is the most correlated one and its orbital energy is even lower than $d_{z^2}$. Such a orbital order flip has also been reported in recent GW+DMFT study \cite{Peto2020}.

\section{Electronic structure with f-electrons} 

\begin{figure*}
	\subfigure[Pristine]{\includegraphics[width=8cm]{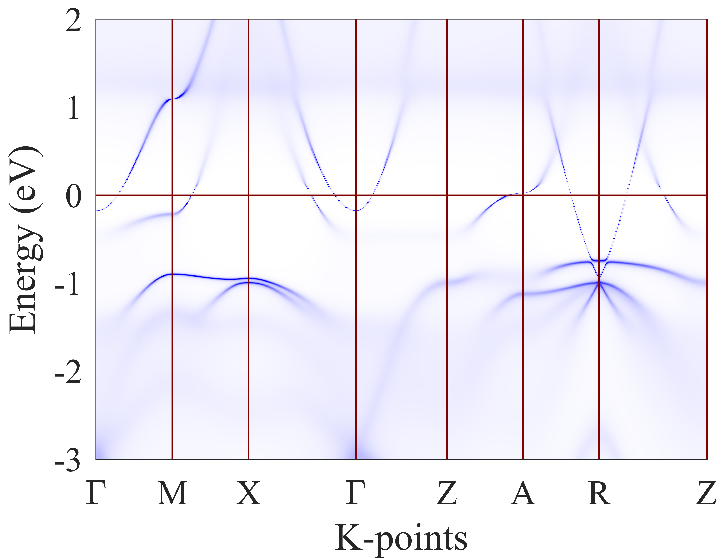}}
	\subfigure[Doped]{\includegraphics[width=8cm]{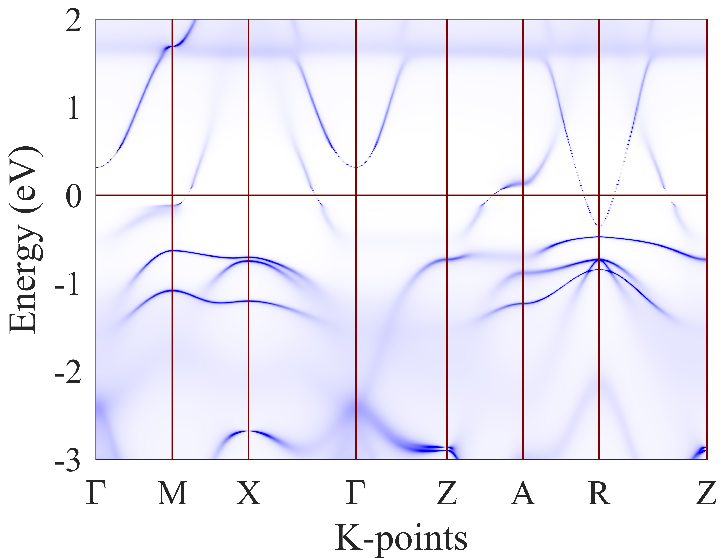}}
	\caption{$\mathbf{k}$-resolved spectral function from DFT+DMFT calculations, by including f-elections.\label{sfig:dmft_spec}}
\end{figure*}

 We show the results from calculation explicitly considering Nd-4$f$ electrons in Fig. \ref{sfig:dmft_spec}. These calculations are done with $U_f=6.0$ eV, $J_f=0.7$ eV on Nd-4$f$ orbitals and $U_d=5.0$ eV, $J_d=0.8$ eV on Ni-3$d$ orbitals. The $\mathbf{k}$-resolved spectral functions are almost the same with the calculations without f-electrons (as shown in Fig. \ref{fig:dmft_spec}). In addition, the 3$d$-orbital occupation and effective masses are also qualitatively the same. Since the calculations are extremely expensive due to Nd-4$f$ orbitals, we have only calculated pristine and 0.2 hole-doped compounds. 
	
	\begin{table}
		\caption{In a calculation with f-electrons, 3d-orbital occupations and effective masses in both pristine and 0.2-hole doped compounds. The numbers within the brackets are the effective mass of respective orbitals. \label{stab:occf}.}
		\begin{tabular}{c|c|c|c|c}
			\hline\hline
			doping  & $d_{x^2-y^2}$ & $d_{z^2}$ & $d_{zx/zy}$ & $d_{xy}$  \\
			\hline
			0.0 & 1.271 (2.30) & 1.631 (1.23) & 3.775 (1.21) & 1.942 (1.23) \\
			0.2 & 1.236 (2.28) & 1.628 (1.24) & 3.806 (1.23) & 1.945 (1.24)\\
			\hline\hline
		\end{tabular}
	\end{table}

\end{document}